\documentclass[12pt]{article}

\usepackage{cite}

\author{Yu.~M.~Zinoviev
       \thanks{E-mail address: Yurii.Zinoviev@ihep.ru} \\[0.5cm]
        {\it Institute for High Energy Physics} \\
        {\it of National Research Center "Kurchatov Institute"} \\
        {\it Protvino, Moscow Region, 142280, Russia}}

\title{Partially massless spin 2 and supersymmetry}

\date{}

\begin{document}

\maketitle

\begin{abstract}
The very existence of partially massless spin 2 supermultiplet tell us
that partially massless spin 2 has two natural superpartners: massless
spin 3/2 and massive spin 3/2 with some special value of mass. As for
any pair of fields connected by global supertransformations there are
two natural questions: existence of the self-interaction and 
possibility to make supertransformations to be local by switching
their interaction with massless spin 3/2 gravitino. At first, we
consider a self-interaction for the partially massless spin 2 and
massive spin 3/2 which may be considered as the first approximation to
partially massless supergravity and  provide a direct construction of
the minimal (i.e. having no more than one derivative) vertex which
resembles usual supergravity. Then we consider localization of global
supersymmetry which connects partially massless spin 2 with its two
possible superpartners --- massless spin 3/2 and massive with special
mass value. For the first case we also managed to construct a minimal
vertex having no more that one derivative. Again this vertex can be
considered as a part of what can be called partially massless $N=2$
supergravity. As for the second case, the corresponding vertex does
exist but it has higher derivative terms.
\end{abstract}

\thispagestyle{empty}
\newpage
\setcounter{page}{1}

\section{Introduction}

Partially massless fields \cite{DW01,DW01a,DW01c,Zin01,Met06,Met22a}
occupy an intermediate position between massless and massive fields so
their study can gives us some useful experience. Till now partially
massless spin 2 attracted the most attention due to the possibility to
construct something like partially massless gravity\footnote{On
the connection of partially massless spin 2 and conformal gravity see
\cite{DJW12,DJW13,HSS13}.}. The cubic vertex describing
self-interaction of this field does exist \cite{Zin06,Zin14} but all
attempts to go beyond the cubic level mostly  lead to the no-go
statements  
\cite{RR12,HSS12,HSS12d,DSW13,RHRT13,JLT14,JR15,GHJMR16,BDGT19,JMP19,ST21,BGPT24}.

The partially massless fields are unitary in de Sitter space and 
non-unitary  in anti de Sitter space and thus they hardly were
associated with supersymmetry. But it appeared that supermultiplets
with partially massless fields (and sometimes with massless ones)
really exist \cite{GHR18,BKhSZ19a,BGHR21}. One of the simplest but
interesting examples is the partially massless spin 2 supermultiplet
which for the first time was constructed in \cite{GHR18}. It tell us
that partially massless spin 2 has two natural superpartners: massless
spin 3/2 and massive spin 3/2 with some special value of
mass\footnote{One more possibility is the partially massless spin 5/2
but this case deserves separate study.}. As for any pair of fields
connected by global supertransformations there are two natural
questions: existence of the self-interaction and  possibility to make
supertransformations to be local by switching their interaction with
massless spin 3/2 gravitino. And these are the aims of the current
work.

Our paper is organized as follows. As in the case of massive
supermultiplets \cite{BKhSZ19,Zin24} for the construction of partially
massless ones and investigation of their interaction it is most
effective to use the gauge invariant description of massive and
partially massless fields \cite{Zin08b,PV10,KhZ19}. Thus in Section 2
we provide all necessary kinematical information for the description
of partially massless spin 2, massive and massless spin 3/2 fields.
In Section 3 we consider a self-interaction for the partially massless
spin 2 and massive spin 3/2 (see also recent work \cite{BLT23}) which
may be considered as the first approximation to partially massless
supergravity. Our general analysis shows that there exist two vertices
for arbitrary value of spin 3/2 mass and one more vertex for its
special value. Then we provide a direct construction of the minimal
(i.e. having no more than one derivative) vertex which resembles usual
supergravity. Recall that we already tried to obtain such vertex as a
partially massless limit for the vertex describing interaction of
massive spin 2 and spin 3/2 fields \cite{Zin18,Zin18a} and it appeared
that for arbitrary values of spin 3/2 mass such limit is singular.
Now we understand that it exists only for special mass value. 

Two other sections devoted to the localization of global supersymmetry
which connects partially massless spin 2 with its two possible
superpartners --- massless spin 3/2 and massive with special mass
value. Let us stress that all our investigations show that for this
purpose one have to introduce an external massless gravitino. It
means that if we keep all for members of the partially massless spin 2
supermultiplet, this will lead us to something like partially
massless bi-supergravity and not to the partially massless $N=2$
supergravity. To find supertransformations and fix ambiguity related
with field redefinitions, we use the procedure (which we already used
for the massive supermultiplets in \cite{KhZ20,Zin24}) based on the
deformation of the unfolded equations in the presence of external
massless gravitino. 

Section 4 devoted to the system partially massless spin 2 and two
massless spin 3/2. Here we also managed to construct a minimal vertex
having no more that one derivative. Again this vertex can be
considered as a part of what can be called partially massless $N=2$
supergravity. Section 5 contains our result on the second case. Here
the non-trivial cubic is also constructed but it has higher derivative
terms.

\section{Kinematics}

In this section we provide all necessary kinematical information on
the gauge invariant description for partially massless spin 2 and
massive spin 3/2 (see \cite{KhZ19} for arbitrary spins and Appendix
for some technical details).

\subsection{Partially massless spin 2}

In the frame-like multispinor formalism we need one-forms 
$\Omega^{\alpha(2)} + h.c.$, $H^{\alpha\dot\alpha}$,
$A$ and zero-form $B^{\alpha(2)} + h.c.$. The free Lagrangian 
(four-form in our framework) looks like:
\begin{eqnarray}
{\cal L}_0 &=& \Omega^{\alpha\beta} E_\beta{}^\gamma
\Omega_{\alpha\gamma} + \Omega^{\alpha\beta} e_\beta{}^{\dot\alpha}
D H_{\alpha\dot\alpha} + 4 E B_{\alpha(2)} B^{\alpha(2)} + 2 
E_{\alpha(2)} B^{\alpha(2)} D A \nonumber \\
 && - 2m E_{\alpha(2)} \Omega^{\alpha(2)} A + 4m B^{\alpha\beta}
E_\beta{}^{\dot\alpha} H_{\alpha\dot\alpha} + h.c. 
\end{eqnarray}
It is invariant under the following gauge transformations
\begin{eqnarray}
\delta \Omega^{\alpha(2)} &=& D \eta^{\alpha(2)}, \qquad 
\delta H^{\alpha\dot\alpha} = D \xi^{\alpha\dot\alpha} + 
e_\beta{}^{\dot\alpha} \eta^{\alpha\beta} + e^\alpha{}_{\dot\beta}
\eta^{\dot\alpha\dot\beta} + m e^{\alpha\dot\alpha} \xi, \nonumber \\
\delta B^{\alpha(2)} &=& - m \eta^{\alpha(2)}, \qquad
\delta A = D \xi + m e^{\alpha\dot\alpha} \xi_{\alpha\dot\alpha}, 
\end{eqnarray}
where
$$
m^2 = - 2\lambda^2.
$$
One of the nice features of the frame-like formalism is the fact that
each field (both physical as well as auxiliary) has its own gauge
invariant object (curvature):
\begin{eqnarray}
{\cal R}^{\alpha(2)} &=& D \Omega^{\alpha(2)} + m E^\alpha{}_\beta
B^{\alpha\beta} \nonumber \\
{\cal T}^{\alpha\dot\alpha} &=& D H^{\alpha\dot\alpha} + 
e_\beta{}^{\dot\alpha} \Omega^{\alpha\beta} + e^\alpha{}_{\dot\beta}
\Omega^{\dot\alpha\dot\beta} + m e^{\alpha\dot\alpha} A \\
{\cal A} &=& D A + 2(E^{\alpha(2)} B_{\alpha(2)} + E^{\dot\alpha(2)}
B_{\dot\alpha(2)}) + m e^{\alpha\dot\alpha} H_{\alpha\dot\alpha}
\nonumber \\
{\cal B}^{\alpha(2)} &=& D B^{\alpha(2)} + m \Omega^{\alpha(2)}
\nonumber
\end{eqnarray}
As it is rather common in gravity and supergravity (and in the 
frame-like formalism in general) we use an analogue of the so-called
torsion zero condition:
\begin{equation}
{\cal T}^{\alpha\dot\alpha} \approx 0, \qquad {\cal A} \approx 0.
\end{equation}
In this case the remaining curvatures satisfy the following
differential identities:
\begin{equation}
D {\cal R}^{\alpha(2)} \approx m E^\alpha{}_\beta {\cal
B}^{\alpha\beta}, \qquad D {\cal B}^{\alpha\beta} \approx m 
{\cal R}^{\alpha\beta}
\end{equation}
as well as algebraic ones
\begin{equation}
e_\beta{}^{\dot\alpha} {\cal R}^{\alpha\beta} + e^\alpha{}_{\dot\beta}
{\cal R}^{\dot\alpha\dot\beta} \approx 0, \qquad
E_{\alpha(2)} {\cal B}^{\alpha(2)} + E_{\dot\alpha(2)}
{\cal B}^{\dot\alpha(2)} \approx 0. 
\end{equation}
The variation of the free Lagrangian under arbitrary physical field
variations can be written as:
\begin{equation}
\delta {\cal L}_0 = - {\cal R}^{\alpha\beta} e_\beta{}^{\dot\alpha}
\delta H_{\alpha\dot\alpha} - 2 E_{\alpha(2)} {\cal B}^{\alpha(2)}
\delta A + h.c. 
\end{equation}

Now let us turn to the unfolded equations \cite{KhZ19,KhZ20}. For the
gauge sector (i.e. sector of gauge one-forms and Stueckelberg
zero-form) we have
\begin{eqnarray}
0 &=& D \Omega^{\alpha(2)} + m E^\alpha{}_\beta B^{\alpha\beta}
+ E_{\beta(2)} W^{\alpha(2)\beta(2)} \nonumber \\
0 &=& D H^{\alpha\dot\alpha} + e_\beta{}^{\dot\alpha}
\Omega^{\alpha\beta} + e^\alpha{}_{\dot\beta}
\Omega^{\dot\alpha\dot\beta} + m e^{\alpha\dot\alpha} A \\
0 &=& D A + 2 (E_{\alpha(2)} \Omega^{\alpha(2)} + E_{\dot\alpha(2)}
\Omega^{\dot\alpha(2)}) + m e_{\alpha\dot\alpha} H^{\alpha\dot\alpha}
\nonumber \\
0 &=& D B^{\alpha(2)} + m \Omega^{\alpha(2)} + e_{\beta\dot\alpha}
B^{\alpha(2)\beta\dot\alpha} \nonumber
\end{eqnarray}
Here $W^{\alpha(4)}$ and $B^{\alpha(3)\dot\alpha}$ are the first
representatives of two infinite chains of gauge invariant zero-forms.
Their unfolded equations are $(k \ge 0)$:
\begin{eqnarray}
0 &=& D W^{\alpha(4+k)\dot\alpha(k)} + e_{\beta\dot\beta}
W^{\alpha(4+k)\beta\dot\alpha(k)\dot\beta} + a_{1,k}
e^\alpha{}_{\dot\beta} B^{\alpha(3+k)\dot\alpha(k)\dot\beta}
+ b_{1,k} e^{\alpha\dot\alpha} W^{\alpha(3+k)\dot\alpha(k-1)}
\nonumber \\
0 &=& D B^{\alpha(3+k)\dot\alpha(k+1)} + e_{\beta\dot\beta}
B^{\alpha(3+k)\beta\dot\alpha(k+1)\dot\beta} + a_{2,k}
e_\beta{}^{\dot\alpha} W^{\alpha(3+k)\beta\dot\alpha(k)} 
+ b_{2,k} e^{\alpha\dot\alpha} B^{\alpha(2+k)\dot\alpha(k)} 
\end{eqnarray}
where
$$
a_{1,k} = \frac{4m}{(k+4)(k+5)}, \qquad
a_{2,k} = \frac{m}{(k+1)(k+2)},
$$
$$
b_{1,k} = \frac{(k+2)(k+3)}{(k+1)(k+4)}\lambda^2, \qquad
b_{2,k} = \frac{k(k+5)}{(k+1)(k+4)}\lambda^2.
$$

\subsection{Massive spin 3/2}

In this case we use one-form $\Phi^\alpha + h.c.$ and zero-form 
$\phi^\alpha + h.c.$. The free Lagrangian looks like:
\begin{eqnarray}
{\cal L}_0 &=& - \Phi_\alpha e^\alpha{}_{\dot\alpha} D
\Phi^{\dot\alpha} - \phi_\alpha E^\alpha{}_{\dot\alpha} D
\phi^{\dot\alpha} \nonumber \\
 && - M \Phi_\alpha E^\alpha{}_\beta \Phi^\beta 
 + a_0 \Phi_\alpha E^\alpha{}_{\dot\alpha} 
 + M E \phi_\alpha \phi^\alpha + h.c. 
\end{eqnarray}
It is invariant under the following gauge transformations
\begin{equation}
\delta \Phi^\alpha = D \rho^\alpha + M 
e^\alpha{}_{\dot\alpha} \rho^{\dot\alpha}, \qquad
\delta \phi^\alpha = a_0 \rho^\alpha,
\end{equation}
where
$$
M^2 = m^2 + \lambda^2, \qquad a_0{}^2 = 6m^2.
$$
Corresponding gauge invariant curvatures
\begin{eqnarray}
{\cal F}^\alpha &=& D \Phi^\alpha + M e^\alpha{}_{\dot\alpha}
\Phi^{\dot\alpha} - \frac{a_0}{3} E^\alpha{}_\beta \phi^\beta
\nonumber \\
{\cal C}^\alpha &=& D \phi^\alpha - a_0 \Phi^\alpha + M 
e^\alpha{}_{\dot\alpha} \phi^{\dot\alpha} 
\end{eqnarray}
satisfy the following differential identities
\begin{eqnarray}
D {\cal F}^\alpha &=& - M e^\alpha{}_{\dot\alpha} 
{\cal F}^{\dot\alpha} - \frac{a_0}{3} E^\alpha{}_\beta {\cal C}^\beta
\nonumber \\
D {\cal C}^\alpha &=& - a_0 {\cal F}^\alpha - M 
e^\alpha{}_{\dot\alpha} {\cal C}^{\dot\alpha}
\end{eqnarray}
Variation of the free Lagrangian under the arbitrary field variations
can be written as
\begin{equation}
\delta {\cal L}_0 = - {\cal F}_{\dot\alpha} e_\alpha{}^{\dot\alpha}
\delta \Phi^\alpha - {\cal C}_{\dot\alpha} E_\alpha{}^{\dot\alpha}
\delta \phi^\alpha + h.c.
\end{equation}
{\bf Special case $M = 0$} In this case the free Lagrangian has the
form
\begin{equation}
{\cal L}_0 = - \Phi_\alpha e^\alpha{}_{\dot\alpha} D \Phi^{\dot\alpha}
- \phi_\alpha E^\alpha{}_{\dot\alpha} D \phi^{\dot\alpha} + a_0
\Phi_\alpha E^\alpha{}_{\dot\alpha} \phi^{\dot\alpha} + h.c. 
\end{equation}
while the gauge transformations now 
\begin{equation}
\delta \Phi^\alpha = D \rho^\alpha, \qquad
\delta \phi^\alpha = a_0 \rho^\alpha.
\end{equation}
Here
$$
m^2 = - \lambda^2.
$$
The gauge invariant curvatures now look like
$$
{\cal F}^\alpha = D \Phi^\alpha - \frac{a_0}{3} E^\alpha{}_\beta
\phi^\beta, \qquad {\cal C}^\alpha = D \phi^\alpha - a_0 \Phi^\alpha 
$$
while the variation of the free Lagrangian is the same as before.

Unfolded equations for the gauge sector are simply
\begin{eqnarray}
0 &=& D \Phi^\alpha - \frac{a_0}{3} E^\alpha{}_\beta \phi^\beta
+ E_{\beta(2)} Y^{\alpha\beta(2)} \nonumber \\
0 &=& D \phi^\alpha - a_0 \Phi^\alpha + e_{\beta\dot\alpha}
\phi^{\alpha\beta\dot\alpha} 
\end{eqnarray}
where $Y^{\alpha(3)}$ and $\phi^{\alpha(2)\dot\alpha}$ are the first
representatives of the two chains of the gauge invariant zero-forms.
In turn, for these gauge invariant zero-forms we have $(k \ge 0)$:
\begin{eqnarray}
0 &=& D Y^{\alpha(3+k)\dot\alpha(k)} + e_{\beta\dot\beta}
Y^{\alpha(3+k)\beta\dot\alpha(k)\dot\beta} + c_{1,k} 
e^\alpha{}_{\dot\beta} \phi^{\alpha(2+k)\dot\alpha(k)\dot\beta}
+ d_{1,k} e^{\alpha\dot\alpha} Y^{\alpha(2+k)\dot\alpha(k-1)}
\nonumber \\
0 &=& D \phi^{\alpha(2+k)\dot\alpha(k+1)} + e_{\beta\dot\beta}
\phi^{\alpha(2+k)\beta\dot\alpha(k+1)\dot\beta} + c_{2,k} 
e_\beta{}^{\dot\alpha} Y^{\alpha(2+k)\beta\dot\alpha(k)} + d_{2.k}
e^{\alpha\dot\alpha} \phi^{\alpha(1+k)\dot\alpha(k)} 
\end{eqnarray}
$$
c_{1,k} = - \frac{a_0}{2(k+3)(k+4)}, \qquad
c_{2,k} = - \frac{a_0}{(k+1)(k+2)},
$$
$$
d_{1,k} = \frac{(k+2)^2}{(k+1)(k+3)}\lambda^2, \qquad
d_{2,k} = \frac{k(k+4)}{(k+1)(k+3)}\lambda^2.
$$

\subsection{Massless spin 3/2}

We need only one-form $\Phi^\alpha + h.c.$. The free Lagrangian looks
like
\begin{equation}
{\cal L}_0 = - \Phi_\alpha e^\alpha{}_{\dot\alpha} D \Phi^{\dot\alpha}
\mp \lambda \Phi_\alpha E^\alpha{}_\beta \Phi^\beta + h.c.
\end{equation}
Note the two possible signs of the $\lambda$-term that will be
important later  on. Gauge transformations leaving this Lagrangian
invariant now
\begin{equation}
\delta \Phi^\alpha = D \rho^\alpha \pm \lambda
e^\alpha{}_{\dot\alpha} \rho^{\dot\alpha}
\end{equation}
while the gauge invariant curvature looks like
\begin{equation}
{\cal F}^\alpha = D \Phi^\alpha \pm \lambda 
e^\alpha{}_{\dot\alpha} \Phi^{\dot\alpha}.
\end{equation}
It satisfies an identity
\begin{equation}
D {\cal F}^\alpha = \mp \lambda e^\alpha{}_{\dot\alpha}
{\cal F}^{\dot\alpha}. 
\end{equation}
Variation of the free Lagrangian has the form
\begin{equation}
\delta {\cal L} = - {\cal F}_{\dot\alpha} e_\alpha{}^{\dot\alpha}
\delta \Phi^\alpha + h.c. 
\end{equation}
Unfolded equations for gauge fields
\begin{eqnarray}
0 &=& D \Phi^\alpha \pm \lambda e^\alpha{}_{\dot\alpha}
\Phi^{\dot\alpha} + E_{\beta(2)} Y^{\alpha\beta(2)} \nonumber \\
0 &=& D \Phi^{\dot\alpha} \pm \lambda e_\alpha{}^{\dot\alpha}
\Phi^\alpha + E_{\dot\beta(2)} Y^{\dot\alpha\dot\beta(2)}
\end{eqnarray}
while equations for the gauge invariant zero-forms look like 
$(k \ge 0)$:
\begin{equation}
0 = D Y^{\alpha(3+k)\dot\alpha(k)} + e_{\beta\dot\beta}
Y^{\alpha(3+k)\beta\dot\alpha(k)\dot\beta} + \lambda^2
e^{\alpha\dot\alpha} Y^{\alpha(2+k)\dot\alpha(k-1)}. 
\end{equation}

\section{Partially massless spin 2 and massive spin 3/2 ---
self interaction}

\subsection{General analysis}

Working with the gauge invariant formulation for massive and partially
massless fields one faces the problem of field redefinitions due to
the presence of Stueckelberg fields \cite{BDGT18,Zin24a}. In this case
(as in general for all cubic vertices with two or three massive
fields) we have enough field redefinitions to bring the vertex into
abelian form. We use this fact to classify possible vertices. First of
all note that using the gauge invariant curvatures one can write three
trivially gauge invariant cubic vertices which do not vanish on-shell:
\begin{equation}
{\cal L}_1 = h_1 {\cal R}^{\alpha\beta} {\cal C}_\alpha {\cal C}_\beta
+ h_2 {\cal B}^{\alpha\beta} {\cal F}_\alpha {\cal C}_\beta + h_3
{\cal B}^{\alpha\beta} e_\alpha{}^{\dot\alpha} {\cal C}_{\dot\alpha}
{\cal C}_\beta + h.c. 
\end{equation}
However they are not independent due to an identity (here and in what
follows $\approx$ means "up to total derivative and terms which vanish
on-shell")
$$
0 \approx D [ {\cal B}^{\alpha\beta} {\cal C}_\alpha {\cal C}_\beta]
= m_0 {\cal R}^{\alpha\beta} {\cal C}_\alpha {\cal C}_\beta + 2
{\cal B}^{\alpha\beta} [ a_0 {\cal F}_\alpha + M 
e_\alpha{}^{\dot\alpha} {\cal C}_{\dot\alpha}] {\cal C}_\beta. 
$$
We choose the vertices with the coefficients $h_{1,3}$ as independent
ones. Now let us consider the most general ansatz for the abelian
vertices:
\begin{eqnarray}
{\cal L}_a &=& g_1 B^{\alpha\beta} {\cal F}_\alpha 
e_\beta{}^{\dot\alpha} {\cal C}_{\dot\alpha} + g_2 B^{\alpha\beta}
e_\alpha{}^{\dot\alpha} {\cal F}_{\dot\alpha} {\cal C}_\beta 
 + g_3 B^{\alpha\gamma} E^\beta{}_{\gamma} {\cal C}_\alpha
{\cal C}_\beta \nonumber \\
 && + g_4 H^{\alpha\dot\alpha} {\cal F}_\alpha {\cal C}_{\dot\alpha}
 + g_5 H^{\alpha\dot\alpha} e^\beta{}_{\dot\alpha} {\cal C}_\alpha
{\cal C}_\beta + g_6 A {\cal F}^\alpha {\cal C}_\alpha + h.c. 
\end{eqnarray}
It appeared that for $M = 0$ all variations under the gauge
transformations vanish on-shell (i.e. can be compensated by the
appropriate corrections) for the combination
\begin{equation}
{\cal L}_a = g_1 B^{\alpha\beta} {\cal F}_\alpha 
e_\beta{}^{\dot\alpha} {\cal C}_{\dot\alpha} + g_4
H^{\alpha\dot\alpha} {\cal F}_\alpha {\cal C}_{\dot\alpha} + h.c.
\end{equation}
where
$$
mg_1 = g_4.
$$
Thus we have three possible vertices: two for arbitrary spin 3/2 mass
and one for $M = 0$ only. It is instructive to consider them in the
unitary gauge (where Stueckelberg zero-forms are set to zero
$B^{\alpha(2)} = 0$, $\phi^\alpha = 0$). We obtain
\begin{equation}
{\cal L}_{u.g.} = a_0{}^2h_1 D \Omega^{\alpha\beta} \Phi_\alpha 
\Phi_\beta + ma_0{}^2h_3 \Omega^{\alpha\beta} e_\alpha{}^{\dot\alpha}
\Phi_{\dot\alpha} \Phi_\beta - a_0g_4 H^{\alpha\dot\alpha}
D \Phi_\alpha \Phi_{\dot\alpha} + h.c. 
\end{equation}
So the first vertex has two derivatives while the second and the third
vertices only one. Moreover, a particular combination of these two can
mimic standard minimal interactions. Our aim in this section is to
construct such minimal vertex (i.e. having no more that one
derivative) using the same constructive approach as in
\cite{Zin18,Zin18a}.

\subsection{Terms with one derivative}

Let us introduce the following ansatz for such terms
\begin{eqnarray}
{\cal L}_{11} &=& c_1 \Omega^{\alpha\beta} \Phi_\alpha 
e_\beta{}^{\dot\alpha} \Phi_{\dot\alpha} + c_2 H^{\alpha\dot\alpha}
{\cal F}_\alpha \Phi_{\dot\alpha} +  c_3 E_\beta{}^{\dot\alpha}
B^{\alpha\beta} \Phi_\alpha \phi_{\dot\alpha} \nonumber \\
 && + c_4 \Omega^{\alpha\beta} E_\beta{}^{\dot\alpha} \phi_\alpha
\phi_{\dot\alpha} + c_5 [ E^\alpha{}_\beta H^{\beta\dot\alpha} -
E^{\dot\alpha}{}_{\dot\beta} H^{\alpha\dot\beta}] \phi_\alpha
{\cal C}_{\dot\alpha} + h.c. 
\end{eqnarray} \and consider its variations under all gauge
transformations.\\
{\bf Lorentz transformations} Variations produce
\begin{eqnarray*}
\delta {\cal L}_{11} &=& - c_1 \eta^{\alpha\beta} 
e_\beta{}^{\dot\alpha} [{\cal F}_\alpha \Phi_{\dot\alpha} -
\Phi_\alpha {\cal F}_{\dot\alpha} + c_2 \eta^{\alpha\beta}
e_\beta{}^{\dot\alpha} [ {\cal F}_\alpha \Phi_{\dot\alpha} +
\Phi_\alpha {\cal F}_{\dot\alpha}] \\ 
 && + \eta^{\alpha\beta} E_\beta{}^{\dot\alpha} [ c_4
({\cal C}_\alpha \phi_{\dot\alpha} + \phi_\alpha 
{\cal C}_{\dot\alpha}) - 2c_5 ({\cal C}_\alpha \phi_{\dot\alpha} -
\phi_\alpha {\cal C}_{\dot\alpha}) ] \\
 && + \frac{a_0}{3}c_1 \eta^{\alpha\beta} E_\beta{}^{\dot\alpha}
[ 3 \Phi_\alpha \phi_{\dot\alpha} + \Phi_{\dot\alpha} \phi_\alpha]
- mc_3 \eta^{\alpha\beta} E_\beta{}^{\dot\alpha} \Phi_\alpha
\phi_{\dot\alpha} \\
 && + a_0c_4 \eta^{\alpha\beta} E_\beta{}^{\dot\alpha} (\Phi_\alpha
\phi_{\dot\alpha} - \Phi_{\dot\alpha} \phi_\alpha)
\end{eqnarray*}
We put
$$
c_2 = c_1, \qquad c_5 = \frac{c_4}{2}
$$
and obtain
\begin{eqnarray}
\delta {\cal L}_{11} &=& - 2c_1 e_\alpha{}^{\dot\alpha} 
{\cal F}_{\dot\alpha} \eta^{\alpha\beta} \Phi_\beta - 2c_4
E_\alpha{}^{\dot\alpha} {\cal C}_{\dot\alpha} \eta^{\alpha\beta}
\phi_\beta \nonumber \\
 && + \frac{a_0}{3}c_1 \eta^{\alpha\beta} E_\beta{}^{\dot\alpha} 
[ 3 \Phi_\alpha \phi_{\dot\alpha} + \Phi_{\dot\alpha} \phi_\alpha]
- mc_3 \eta^{\alpha\beta} E_\beta{}^{\dot\alpha} \Phi_\alpha
\phi_{\dot\alpha} \nonumber \\
 && + a_0c_4 \eta^{\alpha\beta} E_\beta{}^{\dot\alpha} (\Phi_\alpha
\phi_{\dot\alpha} - \Phi_{\dot\alpha} \phi_\alpha)
\end{eqnarray}
The terms in the first line can be compensated by the corrections
\begin{equation}
\delta \Phi^\alpha = - 2c_1 \eta^{\alpha\beta} \Phi_\beta, \qquad
\delta \phi^\alpha = 2c_4 \eta^{\alpha\beta} \phi_\beta
\end{equation}
while the remaining terms we will take into account later on. \\     
{\bf Supertransformations} Variations produce
\begin{eqnarray}
\delta {\cal L}_{11} &=& - c_1 [ {\cal R}^{\alpha\beta}
e_\beta{}^{\dot\alpha} - {\cal R}^{\dot\alpha\dot\beta}
e^\alpha{}_{\dot\beta}] \Phi_{\dot\alpha} \zeta_\alpha 
+ 2c_1 e_\beta{}^{\dot\alpha} \Omega^{\alpha\beta}
{\cal F}_{\dot\alpha} \zeta_\alpha + mc_1 e^{\alpha\dot\alpha} A
{\cal F}_{\dot\alpha} \zeta_\alpha \nonumber \\
 && + c_3 E_\beta{}^{\dot\alpha} {\cal B}^{\alpha\beta}
\phi_{\dot\alpha} \zeta_\alpha + c_3 E_\beta{}^{\dot\alpha}
B^{\alpha\beta} {\cal C}_{\dot\alpha} \zeta_\alpha \nonumber \\
 && + \frac{a_0}{3}c_1 H^{\alpha\dot\alpha} 
E_{\dot\alpha}{}^{\dot\beta} {\cal C}_{\dot\beta} \zeta_\alpha - 4mc_1
[B^{\alpha\beta} E_\beta{}^{\dot\alpha} + B^{\dot\alpha\dot\beta}
E^\alpha{}_{\dot\beta} ] \Phi_{\dot\alpha} \zeta_\alpha \nonumber \\
 && - \frac{a_0}{3}c_1 [ 3 \Omega^{\alpha\beta} E_\beta{}^{\dot\alpha}
+ \Omega^{\dot\alpha\dot\beta} E^\alpha{}_{\dot\beta}]
\phi_{\dot\alpha} \zeta_\alpha + mc_3 \Omega^{\alpha\beta}
E_\beta{}^{\dot\alpha} \phi_{\dot\alpha} \zeta_\alpha \nonumber \\
 && + a_0c_3 [B^{\alpha\beta} E_\beta{}^{\dot\alpha} +
B^{\dot\alpha\dot\beta} E^\alpha{}_{\dot\beta}] \Phi_{\dot\alpha}
\zeta_\alpha - a_0c_4 [ \Omega^{\alpha\beta} E_\beta{}^{\dot\alpha}
- \Omega^{\dot\alpha\dot\beta} E^\alpha{}_{\dot\beta}]
\phi_{\dot\alpha} \zeta_\alpha \nonumber \\
 && - a_0c_5 [E^\alpha{}_\beta H^{\beta\dot\alpha} - 
E^{\dot\alpha}{}_{\dot\beta} H^{\alpha\dot\beta}] 
{\cal C}_{\dot\alpha} \zeta_\alpha 
\end{eqnarray}
Terms in the first two lines can be compensated by the corrections
\begin{eqnarray}
\delta H^{\alpha\dot\alpha} &=& - c_1 \Phi^{\dot\alpha} \zeta^\alpha,
\qquad \delta A = \frac{m}{2a_0} c_1 e_{\alpha\dot\alpha}
\phi^{\dot\alpha} \zeta^\alpha \nonumber  \\
\delta \Phi^\alpha &=& 2c_1 \Omega^{\alpha\beta} \zeta_\beta
- mc_1 A \zeta^\alpha, \qquad \delta \phi^\alpha = - c_3
B^{\alpha\beta} \zeta_\beta. 
\end{eqnarray}

\subsection{Terms without derivatives}

The only possibility here appeared to be
\begin{equation}
{\cal L}_{10} = d_1(E^\alpha{}_\beta H^{\beta\dot\alpha} +
E^{\dot\alpha}{}_{\dot\beta}H^{\alpha\dot\beta})
(\Phi_\alpha \phi_{\dot\alpha} - \Phi_{\dot\alpha} \phi_\alpha)
\end{equation}
For the Lorentz transformations this gives
\begin{equation}
\delta {\cal L}_{10}= -4d_1 E_\beta{}^{\dot\alpha} \eta^{\alpha\beta}
(\Phi_\alpha \phi_{\dot\alpha} - \Phi_{\dot\alpha} \phi_\alpha)
\end{equation}
while for the supertransformations we obtain
\begin{equation}
\delta {\cal L}_{10} = 4d_1 (\Omega^{\alpha\beta} 
E_\beta{}^{\dot\alpha} - \Omega^{\dot\alpha\dot\beta}
E^\alpha{}_{\dot\beta}) \phi_{\dot\alpha} \zeta_\alpha + d_1
(E^\alpha{}_\beta H^{\beta\dot\alpha} + E^{\dot\alpha}{}_{\dot\beta}
H^{\alpha\dot\beta}) {\cal C}_{\dot\alpha}\zeta_\alpha
\end{equation}
Now collecting all the remaining variations we see that all them
cancel provided
$$
c_3 = \frac{4m}{a_0}c_1, \qquad c_4 = - \frac{c_1}{3}, \qquad
d_1 = - \frac{a_0}{6}c_1.
$$

\subsection{Vertex}

Thus we obtain a minimal (i.e. with no more than one derivative) cubic
vertex:
\begin{eqnarray}
{\cal L}_1 &=& c_1 [\Omega^{\alpha\beta} \Phi_\alpha 
e_\beta{}^{\dot\alpha} \Phi_{\dot\alpha} -
\Omega^{\dot\alpha\dot\beta} \Phi_{\dot\alpha} e^\alpha{}_{\dot\beta}
\Phi_\alpha] + c_1 H^{\alpha\dot\alpha} [ {\cal F}_\alpha
\Phi_{\dot\alpha} - {\cal F}_{\dot\alpha} \Phi_\alpha] \nonumber \\
 && + \frac{4m}{a_0}c_1 [E_\alpha{}^{\dot\alpha} B^{\alpha\beta}
\Phi_\beta \phi_{\dot\alpha} + E^\alpha{}_{\dot\beta}
B^{\dot\alpha\dot\beta} \Phi_{\dot\alpha} \phi_\alpha] \nonumber \\
 && - \frac{c_1}{3} [\Omega^{\alpha\beta} E_\beta{}^{\dot\alpha} -
\Omega^{\dot\alpha\dot\beta} E^\alpha{}_{\dot\beta}]
 \phi_\alpha \phi_{\dot\alpha} - \frac{c_1}{6} [ E^\alpha{}_\beta
H^{\beta\dot\alpha} - E^{\dot\alpha}{}_{\dot\beta}
H^{\alpha\dot\beta}] [\phi_\alpha {\cal C}_{\dot\alpha} 
- {\cal C}_\alpha \phi_{\dot\alpha}] \nonumber \\
 && - \frac{a_0}{6}c_1 (E^\alpha{}_\beta H^{\beta\dot\alpha} +
E^{\dot\alpha}{}_{\dot\beta} H^{\alpha\dot\beta})
(\Phi_\alpha \phi_{\dot\alpha} - \Phi_{\dot\alpha} \phi_\alpha).
\end{eqnarray}
Note that all the terms with $\Omega$ and $H$ fields exactly
correspond to the standard minimal gravitational interaction, i.e.
$$
e \Rightarrow e + H, \qquad D \Rightarrow D + \Omega.
$$
We still have to check the invariance under the remaining gauge
transformations.\\
{\bf Pseudo-translations} Variations produce
\begin{eqnarray}
\delta {\cal L}_{11} &=& - 2c_1 \xi^{\alpha\dot\alpha} {\cal F}_\alpha
{\cal F}_{\dot\alpha} + \frac{c_1}{3} (E^\alpha{}_\beta
\xi^{\beta\dot\alpha} - E^{\dot\alpha}{}_{\dot\beta}
\xi^{\alpha\dot\beta}) {\cal C}_\alpha {\cal C}_{\dot\alpha} \nonumber
\\
 && + \frac{a_0c_1}{3} [E^\alpha{}_\beta \xi^{\beta\dot\alpha} +
E^{\dot\alpha}{}_{\dot\beta} \xi^{\alpha\dot\beta}]
({\cal F}_\alpha \phi_{\dot\alpha} - {\cal F}_{\dot\alpha}
\phi_\alpha) 
\end{eqnarray}
Terms in the first line vanish on-shell while the terms in the second
line can be compensated by correction
\begin{equation}
\delta \Phi^\alpha = - \frac{a_0}{3}c_1 e_{\beta\dot\alpha}
\xi^{\beta\dot\alpha} \phi^\alpha.
\end{equation}
{\bf $\xi$-transformations} They produce
\begin{equation}
\delta {\cal L}_{11} = mc_1 e^{\alpha\dot\alpha} ({\cal F}_\alpha
\Phi_{\dot\alpha} - {\cal F}_{\dot\alpha} \Phi_\alpha) \xi - mc_1
E^{\alpha\dot\alpha} (\phi_\alpha {\cal C}_{\dot\alpha} - 
{\cal C}_\alpha \phi_{\dot\alpha}) \xi
\end{equation}
and these terms can be compensated by
\begin{equation}
\delta \Phi^\alpha = mc_1 \Phi^\alpha \xi, \qquad
\delta \phi^\alpha = mc_1 \phi^\alpha \xi.
\end{equation}
Collecting all the corrections to the gauge transformations we obtain
\begin{eqnarray}
\delta H^{\alpha\dot\alpha}&=& - c_1 \Phi^{\dot\alpha} \zeta^\alpha +
h.c. \qquad \delta A = \frac{m}{2a_0}c_1 e_{\alpha\dot\alpha}
\phi^{\dot\alpha} \zeta^\alpha + h.c.  \nonumber \\
\delta \Phi^\alpha &=& - 2c_1 \eta^{\alpha\beta} \Phi_\beta + 2c_1
\Omega^{\alpha\beta} \zeta_\beta - mc_1 A \zeta^\alpha  - 
\frac{a_0}{3}c_1 e_{\beta\dot\alpha} \xi^{\beta\dot\alpha} \phi^\alpha
+ mc_1 \Phi^\alpha \xi \\
\delta \phi^\alpha &=& - \frac{2c_1}{3} \eta^{\alpha\beta} \phi_\beta 
- \frac{4m}{a_0}c_1 B^{\alpha\beta} \zeta_\beta + mc_1 \phi^\alpha \xi
 \nonumber 
\end{eqnarray}

\section{Partially massless spin 2 and massless spin 3/2}

In the cases with one massive (or partially massless) field and two
massless ones we do have some number of possible field redefinitions
but it is not enough to bring the theory into purely abelian form.
Thus in this case we apply a direct construction of the cubic vertex
trying to keep the number of derivatives as low as possible. Our first
task here is to find global supertransformations leaving the sum of
the free Lagrangians invariant. To solve this task we consider a
consistent deformations of unfolded equations in the presence of the
massless background gravitino $\Psi^\alpha$ satisfying
\begin{equation}
D \Psi^\alpha = - \lambda e^\alpha{}_{\dot\alpha} \Psi^{\dot\alpha}.
\end{equation}

\subsection{Supertransformations for spin 2}

The most general ansatz for deformation of gauge invariant zero-forms
\begin{eqnarray}
0 &=& D W^{\alpha(4+k)\dot\alpha(k)} + e_{\beta\dot\beta}
W^{\alpha(4+k)\beta\dot\alpha(k)\dot\beta} + a_{1,k}
e^\alpha{}_{\dot\beta} B^{\alpha(3+k)\dot\alpha(k)\dot\beta}
+ b_{1,k} e^{\alpha\dot\alpha} W^{\alpha(3+k)\dot\alpha(k-1)}
\nonumber \\
 && + \alpha_{1,k} Y^{\alpha(4+k)\dot\alpha(k)\dot\beta}
\Psi_{\dot\beta} + \beta_{1,k} Y^{\alpha(3+k)\dot\alpha(k)}
\Psi^\alpha    \\
0 &=& D B^{\alpha(3+k)\dot\alpha(k+1)} + e_{\beta\dot\beta}
B^{\alpha(3+k)\beta\dot\alpha(k+1)\dot\beta} + a_{2,k}
e_\beta{}^{\dot\alpha} W^{\alpha(3+k)\beta\dot\alpha(k)} 
+ b_{2,k} e^{\alpha\dot\alpha} B^{\alpha(2+k)\dot\alpha(k)} \nonumber
\\
 && + \alpha_{2,k} Y^{\alpha(3+k)\beta\dot\alpha(k+1)} \Psi_\beta +
\beta_{2,k} Y^{\alpha(3+k)\dot\alpha(k)} \Psi^{\dot\alpha}
\end{eqnarray}
Self-consistency requires
$$
\alpha_{1,k} = \alpha_1, \quad 
\alpha_{2,k} = \alpha_2= - \frac{\lambda}{2m}\alpha_1, \quad
\beta_{1,k} = \frac{(k+2)}{(k+4)}\lambda\alpha_1, \quad
\beta_{2,k} = \frac{(k+5)}{4(k+1)}m \alpha_1.
$$
We proceed with the equations for $\Omega$ and $B$ connected with the
sector of gauge invariant zero-forms:
\begin{eqnarray}
0 &=& D \Omega^{\alpha(2)} + m E^\alpha{}_\beta B^{\alpha\beta}
+ E_{\beta(2)} W^{\alpha(2)\beta(2)} + g_1 e_{\beta\dot\alpha}
Y^{\alpha(2)\beta} \Psi^{\dot\alpha} + g_2 e_{\beta\dot\alpha}
Y^{\alpha(2)\beta\gamma\dot\alpha} \Psi_\gamma \nonumber \\
0 &=& D B^{\alpha(2)} + m \Omega^{\alpha(2)} + e_{\beta\dot\alpha}
B^{\alpha(2)\beta\dot\alpha} + g_3 Y^{\alpha(2)\beta} \Psi_\beta 
\end{eqnarray}
we obtain
$$
g_1 = \alpha_1, \qquad g_2 = 0, \qquad 
g_3 = - \frac{\lambda}{2m}\alpha_1.
$$
At last for physical fields we obtain
\begin{eqnarray}
0 &=& D H^{\alpha\dot\alpha} + e_\beta{}^{\dot\alpha}
\Omega^{\alpha\beta} + e^\alpha{}_{\dot\beta}
\Omega^{\dot\alpha\dot\beta} + m e^{\alpha\dot\alpha} A
 + b_1 (\Phi^\alpha \Psi^{\dot\alpha} + \Phi^{\dot\alpha}
\Psi^\alpha) \nonumber \\
0 &=& D A + 2 (E_{\alpha(2)} B^{\alpha(2)} + E_{\dot\alpha(2)}
B^{\dot\alpha(2)}) + m e_{\alpha\dot\alpha} H^{\alpha\dot\alpha}
+ b_2 (\Phi^\alpha \Psi_\alpha + \Phi^{\dot\alpha} \Psi_{\dot\alpha}) 
\end{eqnarray}
where
\begin{equation}
b_1 = - \alpha_1, \qquad b_2 = \frac{\lambda}{m}\alpha_1.
\end{equation}
Note that the  term with the coefficient $b_2$ is possible only for
the case of two different spin 3/2 fields. Moreover, it appeared
crucial that these two fields have opposite signs of their $\lambda$
-terms. Thus for the global supertransformations of the physical
fields we have
\begin{equation}
\delta H^{\alpha\dot\alpha} = b_1 \Phi^\alpha \zeta^{\dot\alpha} +
h.c., \qquad \delta A = b_2 \Phi^\alpha \zeta_\alpha + h.c.
\end{equation}
where
\begin{equation}
D \zeta^\alpha = - \lambda e^\alpha{}_{\dot\alpha} \zeta^{\dot\alpha}.
\end{equation}

\subsection{Supertransformations for spin 3/2}

Here the most general ansatz for the gauge invariant zero-forms has
the form
\begin{eqnarray}
0 &=& D Y^{\alpha(3+k)\dot\alpha(k)} + e_{\beta\dot\beta}
Y^{\alpha(3+k)\beta\dot\alpha(k)\dot\beta} + \lambda^2
e^{\alpha\dot\alpha} Y^{\alpha(2+k)\dot\alpha(k-1)} \nonumber \\
 && + \gamma_{1,k} W^{\alpha(3+k)\beta\dot\alpha(k)} \Psi_\beta
+ \delta_{1,k} W^{\alpha(3+k)\dot\alpha(k-1)} \Psi^{\dot\alpha}
\qquad \delta_{1,0} = 0 \nonumber \\
 && + \gamma_{2,k} B^{\alpha(3+k)\dot\alpha(k)\dot\beta}
\Psi_{\dot\beta} + \delta_{2,k} B^{\alpha(2+k)\dot\alpha(k)}
\Psi^\alpha \qquad \qquad \delta_{2,0 }= 0 
\end{eqnarray}
Self-consistency  requires:
$$
\gamma_{1,k} = \gamma_1, \quad 
\gamma_{2,k} = \gamma_2 = - \frac{2\lambda}{m}\gamma_1, \quad
\delta_{1,k} = \frac{(k+3)}{(k+1)}\lambda\gamma_1, \quad
\delta_{2,k} = \frac{k}{(k+4)}m\gamma_1.
$$
Equation for the gauge field (note the sign of $\lambda$-term):
\begin{eqnarray}
0 &=& D \Phi^\alpha - \lambda e^\alpha{}_{\dot\alpha}
\Phi^{\dot\alpha} + E_{\beta(2)} Y^{\alpha\beta(2)} \nonumber \\
 && + a_1 \Omega^{\alpha\beta} \Psi_\beta + a_2 H^{\alpha\dot\alpha}
\Psi_{\dot\alpha} + a_3 A \Psi^\alpha + a_4 e^\alpha{}_{\dot\alpha}
B^{\dot\alpha\dot\beta} \Psi^{\dot\beta} + a_5 B^{\alpha\beta}
e_{\beta\dot\alpha} \Psi^{\dot\alpha}. 
\end{eqnarray}
Consistency requires
\begin{equation}
a_1 = - \gamma_1, \quad a_2 = - \lambda a_1, \quad
a_3 = - \frac{m}{2}a_1, \quad a_4 = 0, \quad
 a_5 = \frac{2\lambda}{m}a_1,
\end{equation}
while the global supertransformations look like
\begin{equation}
\delta \Phi^\alpha = a_1 \Omega^{\alpha\beta} \zeta_\beta + a_2
H^{\alpha\dot\alpha} \zeta_{\dot\alpha} + a_3 A \zeta^\alpha + a_5
e_{\beta\dot\alpha} B^{\alpha\beta} \zeta^{\dot\alpha}.
\end{equation}

\subsection{Cubic vertex}

To construct an interaction vertex we begin with the transition from
the global supertransformations to the local ones. Let us consider
$\zeta^\alpha$-transformations for the physical fields
\begin{eqnarray}
\delta H^{\alpha\dot\alpha} &=& b_1 \Phi^{\dot\alpha} \zeta^\alpha,
\qquad \delta A = b_2 \Phi^\alpha \zeta_\alpha \nonumber \\
\delta \Phi^\alpha &=& a_1 \Omega^{\alpha\beta} \zeta_\beta + a_3
A \zeta^\alpha \\
\delta \Phi^{\dot\alpha} &=& a_2 H^{\alpha\dot\alpha} \zeta_\alpha
 + a_5 B^{\dot\alpha\dot\beta} e_{\alpha\dot\beta} \zeta^\alpha
\nonumber
\end{eqnarray}
Bosonic variations of the free Lagrangian produce
\begin{equation}
\delta_B {\cal L}_0 = b_1 \Phi_{\dot\alpha} e_\alpha{}^{\dot\alpha}
{\cal R}^{\alpha\beta} \zeta_\beta + 2b_2 \Phi_\alpha
E_{\dot\alpha(2)} {\cal B}^{\dot\alpha(2)} \zeta^\alpha + h.c. 
\end{equation}
while fermionic variations give
\begin{equation}
\delta_F {\cal L}_0 = - {\cal F}_{\dot\alpha} e_\alpha{}^{\dot\alpha}
[ a_1 \Omega^{\alpha\beta} \zeta_\beta + a_3 A \zeta^\alpha ] 
+ {\cal F}_\alpha e^\alpha{}_{\dot\alpha} [ a_2 H^{\beta\dot\alpha}
\zeta_\beta + a_5 B^{\dot\alpha\dot\beta} e_{\beta\dot\beta}
\zeta^\beta] + h.c.
\end{equation}
Now using explicit expressions for ${\cal F}^\alpha$, integrating
by parts, completing terms with explicit derivatives of the bosonic
fields up to their gauge invariant curvatures and using ''zero torsion
conditions'' we obtain
\begin{eqnarray}
\delta_F {\cal L}_0 &=& a_1 \Phi_{\dot\alpha} e_\alpha{}^{\dot\alpha}
{\cal R}^{\alpha\beta} \zeta_\beta - a_5 \Phi_\alpha E_{\dot\alpha(2)}
{\cal B}^{\dot\alpha(2)} \zeta^\alpha \nonumber \\
 && - a_1 \Phi_{\dot\alpha} e_\alpha{}^{\dot\alpha}
\Omega^{\alpha\beta} [ D \zeta_\beta - \lambda e_\beta{}^{\dot\beta}
\zeta_{\dot\beta}] - a_3 \Phi_{\dot\alpha} e_\alpha{}^{\dot\alpha}
A [ D \zeta^\alpha + \lambda e^\alpha{}_{\dot\beta} \zeta^{\dot\beta}]
\nonumber \\
 && + a_2 \Phi_\alpha e^\alpha{}_{\dot\alpha} H^{\beta\dot\alpha}
[ D \zeta_\beta - \lambda e_\beta{}^{\dot\beta} \zeta_{\dot\beta}]
 - a_5 \Phi_\alpha E_{\dot\alpha(2)} B^{\dot\alpha(2)} [ D
\zeta^\alpha + \lambda e^\alpha{}_{\dot\beta} \zeta_{\dot\beta}] 
\end{eqnarray}
Terms in the first line cancel the similar terms in the bosonic
variations at
\begin{equation}
b_1 = - a_1, \qquad b_2 = - \frac{m}{2\lambda}a_1,
\end{equation}
while the remaining terms have exactly the structure that
can be compensated by the interaction with gravitino. So we obtain the
following candidate for the cubic vertex:
\begin{equation}
{\cal L}_1 =  \Phi_{\dot\alpha} e_\alpha{}^{\dot\alpha} [ a_1
\Omega^{\alpha\beta} \Psi_\beta + a_3 A \Psi^\alpha] - \Phi_\alpha 
[ a_2 e^\alpha{}_{\dot\alpha} H^{\beta\dot\alpha} \Psi_\beta
- a_5 E_{\dot\alpha(2)} B^{\dot\alpha(2)} \Psi^\alpha] + h.c. 
\end{equation}
But it is not the end of the story because we must check the
invariance under all gauge symmetries of partially massless spin 2 as
well as under the second supersymmetry. Let us consider them in turn. 

\noindent
{\bf $\eta^{\alpha(2)}$-transformations} We obtain
\begin{eqnarray}
\delta_\eta {\cal L}_1 &=& - a_1 D \Phi_{\dot\alpha} 
e_\alpha{}^{\dot\alpha} \eta^{\alpha\beta} \Psi_\beta - a_1
\Phi_{\dot\alpha}  e_\alpha{}^{\dot\alpha} \eta^{\alpha\beta} D
\Psi_\beta \nonumber \\
 && - 2a_2 \Phi_\alpha E^\alpha{}_\gamma \eta^{\beta\gamma} \Psi_\beta
 + (a_2 + ma_5) \Phi_{\dot\alpha} E_{\alpha(2)} \eta^{\alpha(2)}
\Psi^{\dot\alpha}
\end{eqnarray}
These terms can not be compensated by any corrections to the 
$\eta$-transformations so we must introduce additional terms into our
vertex. In order not to spoil invariance under the supersymmetry these
terms must contain gauge invariant curvature for gravitino. Let us try
\begin{equation}
\Delta {\cal L}_1 = a_0 \Phi_{\dot\alpha} H^{\alpha\dot\alpha}
\tilde{\cal F}_\alpha + h.c.
\end{equation}
By direct calculations we find that at $a_0 = a_1$ we obtain
\begin{equation}
\delta_\eta ({\cal L}_1 + \Delta {\cal L}_1) = - a_1 
{\cal F}_{\dot\alpha} e_\alpha{}^{\dot\alpha} \eta^{\alpha\beta}
\Psi_\beta - a_1 \Phi_\alpha \eta^{\alpha\beta} e_\beta^{\dot\alpha}
\tilde{\cal F}_{\dot\alpha}
\end{equation}
and these terms can be compensated by 
\begin{equation}
\delta \Phi^\alpha = - a_1 \eta^{\alpha\beta} \Psi_\beta, \qquad
\delta \Psi^\alpha = - a_1 \eta^{\alpha\beta} \Phi_\beta. 
\end{equation}
{\bf $\xi^{\alpha\dot\alpha}$-transformations} Here we get
\begin{equation}
\delta_\xi {\cal L}_1 = a_2 {\cal F}_\alpha e^\alpha{}_{\dot\alpha}
\xi^{\beta\dot\alpha} \Psi_\beta + a_2 \Phi_\alpha 
e^\alpha{}_{\dot\alpha} \xi^{\beta\dot\alpha} \tilde{\cal F}_\beta +
h.c.
\end{equation}
and
\begin{equation}
\delta_\xi (\Delta {\cal L}_1) = a_1 {\cal F}_{\dot\alpha}
\xi^{\alpha\dot\alpha} \tilde{\cal F}_\alpha + \lambda a_1 \Phi_\alpha
(e^\alpha{}_{\dot\alpha} \xi^{\beta\dot\alpha} + 
e^\beta{}_{\dot\alpha} \xi^{\alpha\dot\alpha}) \tilde{\cal F}_\beta +
h.c. 
\end{equation}
Taking into account that
$$
{\cal F}_{\dot\alpha} \xi^{\alpha\dot\alpha} \tilde{\cal F}_\alpha
\approx 0
$$
we obtain
\begin{equation}
\delta ({\cal L}_1 + \Delta {\cal L}_1) = a_2 {\cal F}_\alpha
e^\alpha{}_{\dot\alpha}\xi^{\beta\dot\alpha} \Psi_\beta 
+ a_2 \Phi_\alpha \xi^{\alpha\dot\alpha} e^\beta{}_{\dot\alpha}
\tilde{\cal F}_\beta + h.c. 
\end{equation}
that can be compensated by
\begin{equation}
\delta \Phi^\alpha = - a_2 \xi^{\alpha\dot\alpha} \Psi_{\dot\alpha},
\qquad \delta \Psi^\alpha = - a_2 \xi^{\alpha\dot\alpha}
\Phi_{\dot\alpha}.
\end{equation}
{\bf $\xi$-transformations} Here we get
\begin{eqnarray*}
\delta_\xi {\cal L}_1 &=& a_3 [ {\cal F}_{\dot\alpha}
e^{\alpha\dot\alpha} \Psi_\alpha
+ \Phi_{\dot\alpha} e^{\alpha\dot\alpha} \tilde{\cal F}_\alpha] \xi \\
\delta_\xi (\Delta {\cal L}_1) &=& - 2a_3 \Phi_{\dot\alpha}
e^{\alpha\dot\alpha} \tilde{\cal F}_\alpha \xi + h.c.
\end{eqnarray*}
So we have
\begin{equation}
\delta_\xi ({\cal L}_1 + \Delta{\cal L}_1) = a_3 
[ {\cal F}_{\dot\alpha} e^{\alpha\dot\alpha} \Psi_\alpha -
\Phi_{\dot\alpha} e^{\alpha\dot\alpha} \tilde{\cal F}_\alpha] \xi +
h.c. 
\end{equation}
while appropriate corrections are:
\begin{equation}
\delta \Phi^\alpha = - a_3 \Psi^\alpha \xi, \qquad
\delta \Psi^\alpha = - a_3 \Phi^\alpha \xi. 
\end{equation}
{\bf $\rho^\alpha$-transformations} Similarly to the
supertransformations
\begin{eqnarray}
\delta_\rho {\cal L}_1 &=& - a_1 \rho_\alpha e^\alpha{}_{\dot\alpha}
{\cal R}^{\dot\alpha\dot\beta} \Psi_{\dot\beta} - a_5 \rho_\alpha
E_{\dot\alpha(2)} {\cal B}^{\dot\alpha(2)} \Psi^\alpha \nonumber \\
 && + a_1 \rho_\alpha e^\alpha{}_{\dot\alpha}
\Omega^{\dot\alpha\dot\beta} \tilde{\cal F}_{\dot\beta} + a_2
\rho_\alpha e^\alpha{}_{\dot\alpha} H^{\beta\dot\alpha}
\tilde{\cal F}_\beta \nonumber \\
 && - a_5 \rho_\alpha E_{\dot\alpha(2)} B^{\dot\alpha(2)}
\tilde{\cal F}^\alpha + a_3 \rho_\alpha e^\alpha{}_{\dot\alpha} A
\tilde{\cal F}^{\dot\alpha} 
\end{eqnarray}
\begin{equation}
\delta (\Delta {\cal L}_1) = - a_1 \rho_\alpha 
[ e_\beta{}^{\dot\alpha} \Omega^{\alpha\beta}
+ e^\alpha{}_{\dot\beta} \Omega^{\dot\alpha\dot\beta} + m
e^{\alpha\dot\alpha} A ] \tilde{\cal F}_{\dot\alpha} 
+ \lambda a_1 \rho_\alpha [ e^\alpha{}_{\dot\alpha}
H^{\beta\dot\alpha} + e^\beta{}_{\dot\alpha} H^{\alpha\dot\alpha}]
\tilde{\cal F}_\beta 
\end{equation}
Thus we obtain
\begin{eqnarray*}
\delta_\rho ({\cal L}_1 + \Delta {\cal L}_1) &=&
- a_1 \rho_\alpha e^\alpha{}_{\dot\alpha}
{\cal R}^{\dot\alpha\dot\beta} \Psi_{\dot\beta} - a_5 \rho_\alpha
E_{\dot\alpha(2)} {\cal B}^{\dot\alpha(2)} \Psi^\alpha \\
 && + a_1 \rho_\alpha \Omega^{\alpha\beta} e_\beta{}^{\dot\alpha}
\tilde{\cal F}_{\dot\alpha} + a_2 \rho_\alpha H^{\alpha\dot\alpha}
e^\beta{}_{\dot\alpha} \tilde{\cal F}_\beta 
 - a_3 \rho_\alpha e^\alpha{}_{\dot\alpha} A 
\tilde{\cal F}^{\dot\alpha} - a_5 \rho_\alpha e^\alpha{}_{\dot\alpha}
B^{\dot\alpha\dot\beta} e^\beta{}_{\dot\alpha}
\tilde{\cal F}_\beta 
\end{eqnarray*}
All terms can be compensated by the corrections.

Thus we obtain the following cubic vertex
\begin{eqnarray}
{\cal L}_1 &=& [ a_1 \Phi_{\dot\alpha} e_\beta{}^{\dot\alpha}
\Omega^{\alpha\beta} - a_3 \Phi_{\dot\alpha} e^{\alpha\dot\alpha} A
- a_2 \Phi_\beta e^\beta{}_{\dot\alpha} H^{\alpha\dot\alpha}
 - a_5 \Phi^\alpha E_{\dot\alpha(2)} B^{\dot\alpha(2)}] \Psi_\alpha
\nonumber \\
 && + \frac{a_1}{2} \Phi_{\dot\alpha} H^{\alpha\dot\alpha}
\tilde{\cal F}_\alpha + h.c. 
\end{eqnarray}
while a complete set of gauge transformations has the form:
\begin{eqnarray}
\delta H^{\alpha\dot\alpha} &=& - a_1 [ \Phi^\alpha \zeta^{\dot\alpha}
- \rho^\alpha \Psi^{\dot\alpha} + h.c.], \qquad
\delta A = - \frac{m}{2\lambda}a_1 [ \Phi^\alpha \zeta_\alpha -
\rho^\alpha \Psi_\alpha + h.c. ] \nonumber \\
\delta \Phi^\alpha &=& a_1 \Omega^{\alpha\beta} \zeta_\beta - \lambda
a_1 H^{\alpha\dot\alpha} \zeta_{\dot\alpha} - \frac{m}{2}a_1 A
\zeta^\alpha + \frac{2\lambda}{m}a_1 B^{\alpha\beta}
e_{\beta\dot\alpha} \zeta^{\dot\alpha} \nonumber \\
 && - a_1 \eta^{\alpha\beta} \Psi_\beta + \lambda a_1
\xi^{\alpha\dot\alpha} \Psi_{\dot\alpha} + \frac{m}{2}a_1 \xi
\Psi^\alpha \\
\delta \Psi^\alpha &=& a_1 \Omega^{\alpha\beta} \rho_\beta + \lambda
a_1 H^{\alpha\dot\alpha} \rho_{\dot\alpha} - \frac{m}{2}a_1 A
\rho^\alpha + \frac{2\lambda}{m} a_1 B^{\alpha\beta}
e_{\beta\dot\beta} \rho^{\dot\beta} \nonumber \\
 && - a_1 \eta^{\alpha\beta} \Phi_\beta + \lambda a_1
\xi^{\alpha\dot\alpha} \Phi_{\dot\alpha} + \frac{m}{2}a_1 \xi
\Phi^\alpha \nonumber 
\end{eqnarray}

\section{Partially massless spin 2 and massive spin 3/2 at $M=0$}

\subsection{General analysis for $M \neq 0$}

In this case (one massless field and two massive or partially massless
ones) we have enough field redefinitions to bring the model into
purely abelian form (but in general not to trivially gauge invariant
form). Let us start with the only trivially gauge invariant vertex
that do not vanish on-shell:
\begin{equation}
{\cal L}_1 = h_0 \tilde{\cal F}_\alpha {\cal B}^{\alpha\beta} 
{\cal C}_\beta + h.c. 
\end{equation}
It is equivalent to the following combination of abelian ones
\begin{equation}
\frac{1}{h_0} {\cal L}_1 = \Psi_\alpha [  m {\cal R}^{\alpha\beta}
{\cal C}_\beta + a_0 {\cal B}^{\alpha\beta} {\cal F}_\beta 
- M {\cal B}^{\alpha\beta} e_\beta{}^{\dot\alpha} 
{\cal C}_{\dot\alpha} - \lambda e^\alpha{}_{\dot\alpha} 
{\cal B}^{\dot\alpha\dot\beta} {\cal C}_{\dot\beta}] + h.c.
\end{equation}
Now let us consider general abelian vertex
\begin{equation}
{\cal L}_2 = \Psi_\alpha [ g_1 {\cal R}^{\alpha\beta} {\cal C}_\beta
+ g_2 {\cal B}^{\alpha\beta} {\cal F}_\beta + f_1 
{\cal B}^{\alpha\beta} e_\beta{}^{\dot\alpha} {\cal C}_{\dot\alpha}
+ f_2 e^\alpha{}_{\dot\alpha} {\cal B}^{\dot\alpha\dot\beta}
{\cal C}_{\dot\beta}] + h.c.
\end{equation}
and require it to be supersymmetric. Heavily using differential and
algebraic identities for the curvatures we obtain
\begin{eqnarray}
\delta {\cal L}_2 &=& [ a_0g_1 - m_0g_2  ] \zeta_\alpha 
{\cal R}^{\alpha\beta} {\cal F}_\beta \nonumber \\
 && + [ - (M+\lambda)g_1 - m_0 f_1 - m_0f_2] 
\zeta_\alpha {\cal R}^{\alpha\beta}
e_\beta{}^{\dot\alpha} {\cal C}_{\dot\alpha} \nonumber \\
 && + [ Mg_2 + a_0f_1  ] \zeta_\alpha {\cal B}^{\alpha\beta} 
e_\beta{}^{\dot\alpha}{\cal F}_{\dot\alpha} \nonumber \\
 && + [ \lambda g_2 + a_0f_2  ] \zeta_\alpha e^\alpha{}_{\dot\alpha}
{\cal B}^{\dot\alpha\dot\beta} {\cal F}_{\dot\beta} \nonumber \\
 && + [  - m_0g_1 + 2\lambda f_2 ]
\zeta_\alpha E^\alpha{}_\gamma {\cal B}^{\beta\gamma} {\cal C}_\beta 
\nonumber \\
 && + [ - m_0g_1 + \frac{a_0}{3}g_2 + 2Mf_1 ] 
\zeta_\alpha E^\beta{}_\gamma {\cal B}^{\alpha\gamma} {\cal C}_\beta 
\nonumber \\
 && + [ - \lambda f_1 + Mf_2  ] \zeta_\alpha 
E_{\dot\alpha(2)} {\cal B}^{\dot\alpha(2)} {\cal C}^\alpha 
\end{eqnarray}
First of all it is easy to check that for the parameters
$$
g_1 = m_0, \qquad g_2 = a_0, \qquad f_1 = - M, \qquad f_2 = - \lambda
$$
corresponding to the trivially invariant vertex all these variations
vanish and for $M \ne 0$ it is the only solution. Now taking into
account that terms containing
$$
{\cal R}^{\alpha\beta} e_\beta{}^{\dot\alpha}, \qquad
e_\beta{}^{\dot\alpha} {\cal F}_{\dot\alpha}, \qquad
E_{\alpha(2)} {\cal B}^{\alpha(2)}, \qquad
E_{\dot\alpha(2)} {\cal B}^{\dot\alpha(2)}
$$
can be compensated by the appropriate corrections to the
supertransformations
\begin{equation}
\delta H^{\alpha\dot\alpha} \sim \zeta^\alpha {\cal C}^{\dot\alpha},
\qquad \delta \Phi^\alpha \sim {\cal B}^{\alpha\beta} \zeta_\beta,
\qquad \delta A \sim \zeta_\alpha {\cal C}^\alpha \label{abel}
\end{equation}
we obtain one non-trivial solution at $M=0$, namely
\begin{equation}
{\cal L}_2 = f_1 \Psi_\alpha {\cal B}^{\alpha\beta} 
e_\beta{}^{\dot\alpha} {\cal C}_{\dot\alpha} + h.c. 
\end{equation}
Note that in the unitary gauge $B^{\alpha(2)} = 0$, $\phi^\alpha = 0$
it gives
\begin{equation}
{\cal L} \sim \Psi_\alpha \Omega^{\alpha\beta} e_\beta{}^{\dot\alpha}
\Phi_{\dot\alpha} + h.c. 
\end{equation} 
In what follows we try to reconstruct this vertex using the same
approach as in the previous section.

\subsection{Supertransformations for spin 2}

Ansatz for the gauge invariant zero-forms:
\begin{eqnarray}
0 &=& D W^{\alpha(4+k)\dot\alpha(k)} + e_{\beta\dot\beta}
W^{\alpha(4+k)\beta\dot\alpha(k)\dot\beta} + a_{1,k} 
e^\alpha{}_{\dot\beta} B^{\alpha(3+k)\dot\alpha(k)\dot\beta}
+ b_{1,k} e^{\alpha\dot\alpha} W^{\alpha(3+k)\dot\alpha(k-1)}
\nonumber \\
 && + \alpha_{1,k} Y^{\alpha(4+k)\dot\alpha(k)\dot\beta}
\Psi_{\dot\beta} + \beta_{1,k} Y^{\alpha(3+k)\dot\alpha(k)}
\Psi^\alpha  \\
 && \nonumber \\
0 &=& D B^{\alpha(3+k)\dot\alpha(k+1)} +  e_{\beta\dot\beta}
B^{\alpha(3+k)\beta\dot\alpha(k+1)\dot\beta} + a_{2,k}
e_\beta{}^{\dot\alpha} W^{\alpha(3+k)\beta\dot\alpha(k)} 
 + b_{2,k} e^{\alpha\dot\alpha} B^{\alpha(2+k)\dot\alpha(k)} \nonumber
\\
 && + \alpha_{2,k} Y^{\alpha(3+k)\beta\dot\alpha(k+1)} \Psi_\beta
+ \beta_{2,k} Y^{\alpha(3+k)\dot\alpha(k)} \Psi^{\dot\alpha} \nonumber
\\
 && + \alpha_{3,k} \phi^{\alpha(3+k)\dot\alpha(1+k)\dot\beta}
\Psi_{\dot\beta} + \beta_{3,k} \phi^{\alpha(2+k)\dot\alpha(k+1)}
\Psi^\alpha
\end{eqnarray}
Consistency requires
$$
\alpha_{1,k} = \alpha_1, \qquad 
\alpha_{2,k} = \alpha_2 = - \frac{\lambda}{4m}\alpha_1, \qquad
\alpha_{3,k} = \alpha_3 = - \frac{a_0}{8m}\alpha_1, 
$$
$$
\beta_{1,k} = \frac{(k+3)}{(k+4)}\lambda\alpha_1, \qquad
\beta_{2,k} = \frac{(k+3)(k+5)}{8(k+1)(k+2)}m\alpha_1, \qquad
\beta_{3,k} = - \frac{(k+5)}{(k+4)} \frac{\lambda a_0}{8m}\alpha_1. 
$$
The most general ansatz for $\Omega$ and $B$ fields:
\begin{eqnarray}
0 &=& D \Omega^{\alpha(2)} + m E^\alpha{}_\beta B^{\alpha\beta}
+ E_{\beta(2)} W^{\alpha(2)\beta(2)} \nonumber \\
 && + b_1 \Phi^\alpha \Psi^\alpha + b_2 e^\alpha{}_{\dot\alpha}
\phi^\alpha \Psi^{\dot\alpha} + b_3 e^\alpha{}_{\dot\alpha}
\phi^{\dot\alpha} \Psi^\alpha \nonumber \\
 && + g_1 e_{\beta\dot\alpha}  Y^{\alpha(2)\beta} \Psi^{\dot\alpha} +
g_2 e_{\beta\dot\alpha} \phi^{\alpha(2)\dot\alpha} \Psi^\beta 
+ g_3 e_{\beta\dot\alpha} \phi^{\alpha\beta\dot\alpha} \Psi^\alpha  \\
0 &=& D B^{\alpha(2)} + m \Omega^{\alpha(2)} + e_{\beta\dot\alpha}
B^{\alpha(2)\beta\dot\alpha} \nonumber \\
 && + b_9 \phi^\alpha \Psi^\alpha + g_4 Y^{\alpha(2)\beta} \Psi_\beta
+ g_5 \phi^{\alpha(2)\dot\alpha} \Psi_{\dot\alpha} 
\end{eqnarray}
Consistency with the gauge invariant zero forms sector gives
$$
b_1 = - \frac{\lambda}{2}\alpha_1, \qquad
b_2 = - \frac{a_0}{12}\alpha_1, \qquad 
b_3 = 0, \qquad b_9 = \frac{ma_0}{12\lambda}\alpha_1, 
$$
$$
g_1 = \alpha_1, \qquad g_2 = g_3 = 0, \qquad
g_4 = \alpha_2, \qquad g_5 = \alpha_3. 
$$
At last an ansatz for the physical fields $H$ and $A$:
\begin{eqnarray}
0 &=& D H^{\alpha\dot\alpha} + e_\beta{}^{\dot\alpha}
\Omega^{\alpha\beta} + e^\alpha{}_{\dot\beta}
\Omega^{\dot\alpha\dot\beta} + m e^{\alpha\dot\alpha} A \nonumber \\
 && + b_4 (\Phi^\alpha \Psi^{\dot\alpha} + \Phi^{\dot\alpha}
\Psi^\alpha) + b_5 (e^\alpha{}_{\dot\beta} \phi^{\dot\alpha}
\Psi^{\dot\beta} + e_\beta{}^{\dot\alpha} \phi^\alpha \Psi^\beta)
+ b_6 (e^\alpha{}_{\dot\beta} \phi^{\dot\beta} \Psi^{\dot\alpha}
+ e_\beta{}^{\dot\alpha} \phi^\beta \Psi^\alpha) \\
0 &=& D A + 2 (E_{\alpha(2)} B^{\alpha(2)} + E_{\dot\alpha(2)}
B^{\dot\alpha(2)}) + m e_{\alpha\dot\alpha} H^{\alpha\dot\alpha}
\nonumber \\
 && + b_7 (\Phi^\alpha \Psi_\alpha + \Phi^{\dot\alpha}
\Psi_{\dot\alpha}) + b_8 e_{\alpha\dot\alpha} (\phi^\alpha
\Psi^{\dot\alpha} + \phi^{\dot\alpha} \Psi^\alpha)
\end{eqnarray}
Here we obtain
\begin{equation}
b_4 = - \alpha_1, \qquad
b_5 = b_6 = 0, \qquad
b_7 = - \frac{\lambda}{2m}b_4, \qquad
b_8 = \frac{a_0}{4m}b_4.
\end{equation}
Thus we have the following supertransformations for the physical
fields
\begin{equation}
\delta H^{\alpha\dot\alpha} = b_4 \Phi^{\dot\alpha} \zeta^\alpha +
h.c., \qquad \delta A = b_7 \Phi^\alpha \zeta_\alpha + b_8
e_{\alpha\dot\alpha} \phi^{\dot\alpha} \zeta^\alpha + h.c. 
\end{equation}

\subsection{Supertransformations for spin 3/2}

Ansatz for gauge invariant zero-forms:
\begin{eqnarray}
0 &=& D Y^{\alpha(3+k)\dot\alpha(k)} + e_{\beta\dot\beta} 
Y^{\alpha(3+k)\beta\dot\alpha(k)\dot\beta} + c_{1,k}
e^\alpha{}_{\dot\beta} \phi^{\alpha(2+k)\dot\alpha(k)\dot\beta}
+ d_{1,k} e^{\alpha\dot\alpha} Y^{\alpha(2+k)\dot\alpha(k-1)}
\nonumber \\
 && + \gamma_{1,k} W^{\alpha(3+k)\beta\dot\alpha(k)} \Psi_\beta
+ \delta_{1,k} W^{\alpha(3+k)\dot\alpha(k-1)} \Psi^{\dot\alpha}
\nonumber \\
 && + \gamma_{2,k} B^{\alpha(3+k)\dot\alpha(k)\dot\beta}
\Psi_{\dot\beta} + \delta_{2,k} B^{\alpha(2+k)\dot\alpha(k)}
\Psi^\alpha   \\
0 &=& D \phi^{\alpha(2+k)\dot\alpha(k+1)} + e_{\beta\dot\beta}
\phi^{\alpha(2+k)\beta\dot\alpha(k+1)\dot\beta} + c_{2,k} 
e_\beta{}^{\dot\alpha} Y^{\alpha(2+k)\beta\dot\alpha(k)} 
 + d_{2,k} e^{\alpha\dot\alpha} \phi^{\alpha(1+k)\dot\alpha(k)}
\nonumber \\
 && + \gamma_{3,k} B^{\alpha(2+k)\beta\dot\alpha(k+1)} \Psi_\beta
+ \delta_{3,k} B^{\alpha(2+k)\dot\alpha(k)} \Psi^{\dot\alpha} 
\end{eqnarray}
We obtain
$$
\gamma_{1,k} = \gamma_1, \qquad
\gamma_{2,k} = \gamma_2 = - \frac{\lambda}{m}\gamma_1, \qquad
\gamma_{3,k} = \gamma_3 = - \frac{a_0}{m}\gamma_1,
$$
$$
\delta_{1,k} = \frac{(k+2)}{(k+1)}\lambda\gamma_1, \qquad
\delta_{2,k} = \frac{k(k+2)}{2(k+3)(k+4)}m\gamma_1, \qquad
\delta_{3,k} = - \frac{k}{(k+1)}\frac{\lambda a_0}{m}\gamma_3.
$$
Ansatz for the physical fields:
\begin{eqnarray}
0 &=& D \Phi^\alpha - \frac{a_0}{3} E^\alpha{}_\beta \phi^\beta +
E_{\beta(2)} Y^{\alpha\beta(2)} \nonumber \\
 && + a_1 \Omega^{\alpha\beta} \Psi_\beta + a_2 H^{\alpha\dot\alpha}
\Psi_{\dot\alpha} + a_3 A \Psi^\alpha + a_4 e^\alpha{}_{\dot\beta}
B^{\dot\alpha\dot\beta} \Psi_{\dot\beta} + a_5 e_{\beta\dot\alpha}
B^{\alpha\beta} \Psi^{\dot\alpha} \\
0 &=& D \phi^\alpha - a_0 \Phi^\alpha + e_{\beta\dot\alpha}
\phi^{\alpha\beta\dot\alpha} + a_6 B^{\alpha\beta} \Psi_\beta
\nonumber 
\end{eqnarray}
This leads to 
\begin{equation}
a_2 = a_3 = a_4 = 0, \qquad
a_1 = - \gamma_1, \qquad
a_5 = \frac{\lambda}{m}a_1, \qquad
a_6 = \frac{a_0}{m}a_1.
\end{equation}
 Thus for the supertransformations we get
\begin{equation}
\delta \Phi^\alpha = a_1 \Omega^{\alpha\beta} \zeta_\beta + a_5
e_{\beta\dot\alpha} B^{\alpha\beta} \zeta^{\dot\alpha}, \qquad
\delta \phi^\alpha = - a_6 B^{\alpha\beta} \zeta_\beta .
\end{equation}

\subsection{Cubic vertex}

Let us consider supertransformations for the physical fields
\begin{eqnarray}
\delta H^{\alpha\dot\alpha} &=& b_4 \Phi^{\dot\alpha} \zeta^\alpha,
\qquad \delta A = b_7 \Phi^\alpha \zeta_\alpha + b_8
e_{\alpha\dot\alpha} \phi^{\dot\alpha} \zeta^\alpha \\
\delta \Phi^\alpha &=& a_1 \Omega^{\alpha\beta} \zeta_\beta, \qquad
\delta \Phi^{\dot\alpha} = a_5 B^{\dot\alpha\dot\beta}
e_{\alpha\dot\beta} \zeta^\alpha, \qquad \delta \phi^\alpha = - a_6
B^{\alpha\beta} \zeta_\beta
\end{eqnarray}
Bosonic variations produce
\begin{equation}
\delta_B {\cal L}_0 = b_4 \Phi_{\dot\alpha} e_\alpha{}^{\dot\alpha}
 {\cal R}^{\alpha\beta} \zeta_\beta + 2b_7 \Phi_\alpha
E_{\dot\alpha(2)} {\cal B}^{\dot\alpha(2)} \zeta^\alpha 
+ 4b_8 \phi_{\dot\alpha} E_\alpha{}^{\dot\alpha} 
{\cal B}^{\alpha\beta} \zeta_\beta
\end{equation}
while fermionic variations give
\begin{equation}
\delta_F {\cal L}_0 = - a_1 {\cal F}_\alpha e_\alpha{}^{\dot\alpha}
\Omega^{\alpha\beta} \zeta_\beta - a_5 {\cal F}_\alpha
(E_{\dot\alpha(2)} B^{\dot\alpha(2)}) \zeta^\alpha + a_6 
{\cal C}_{\dot\alpha} E_\alpha{}^{\dot\alpha} B^{\alpha\beta}
\zeta_\beta 
\end{equation}
Using explicit expressions for fermionic curvatures, integrating by
parts and completing derivatives of the bosonic fields up to their
curvatures we obtain
\begin{eqnarray}
\delta_F {\cal L}_0 &=& a_1 \Phi_{\dot\alpha} e_\alpha{}^{\dot\alpha} 
{\cal R}^{\alpha\beta} \zeta_\beta - a_5 \Phi_\alpha E_{\dot\alpha(2)}
{\cal B}^{\dot\alpha(2)} \zeta^\alpha + a_6 \phi_{\dot\alpha} 
E_\alpha{}^{\dot\alpha} {\cal B}^{\alpha\beta} \zeta_\beta \nonumber 
\\
 && - a_1 \Phi_{\dot\alpha} e_\alpha{}^{\dot\alpha}
\Omega^{\alpha\beta} (D \zeta_\beta - \lambda e_\beta{}^{\dot\beta}
\zeta_{\dot\beta}) - a_5 \Phi_\alpha E_{\dot\alpha(2)}
B^{\dot\alpha(2)} (D \zeta^\alpha + \lambda e^\alpha{}_{\dot\beta}
\zeta^{\dot\beta}) \nonumber \\
 && + a_6 \phi_{\dot\alpha} E_\alpha{}^{\dot\alpha} B^{\alpha\beta} 
(D \zeta_\beta - \lambda e_\beta{}^{\dot\beta} \zeta_{\dot\beta})
\end{eqnarray}
Here we use results from the deformation procedure
$$
a_5 = \frac{\lambda}{m}a_1, \qquad a_6 = \frac{a_0}{m}a_1.
$$
The first line cancel the bosonic variations at
$$
b_4 = - a_1, \qquad b_7 = \frac{\lambda}{2m}a_1, \qquad
b_8 = - \frac{a_0}{4m}a_1,
$$
leaving us with the candidate for the cubic vertex
\begin{equation}
{\cal L}_1 = [ a_1 \Phi_{\dot\alpha} e_\alpha{}^{\dot\alpha}
\Omega^{\alpha\beta} - a_5 \Phi^\beta E_{\dot\alpha(2)}
B^{\dot\alpha(2)} - a_6 \phi_{\dot\alpha}  E_\alpha{}^{\dot\alpha}
B^{\alpha\beta}] \Psi_\beta + h.c. 
\end{equation}
But we still have to consider all other gauge transformations. \\
{\bf $\eta^{\alpha(2)}$-transformations} Taking into account
correction
\begin{equation}
\delta_1 \Phi^\alpha = - a_1 \eta^{\alpha\beta} \Psi_\beta
\end{equation}
we obtain
\begin{equation}
\delta_0 {\cal L}_1 + \delta_1 {\cal L}_0 = - a_1 \Phi_{\dot\alpha}
e_\alpha{}^{\dot\alpha} \eta^{\alpha\beta} \tilde{\cal F}_\beta.
\end{equation}
The only possibility we have found is
\begin{equation}
\Delta_1 {\cal L}_1 = - \frac{a_1}{m} \Phi_{\dot\alpha} 
e_\alpha{}^{\dot\alpha} B^{\alpha\beta} \tilde{\cal F}_\beta + h.c. 
\end{equation}
{\bf $\rho^\alpha$-transformations} Here we obtain
\begin{eqnarray*}
\delta {\cal L}_1 &=& a_1 \rho_\alpha e^\alpha{}_{\dot\alpha}
\Omega^{\dot\alpha\dot\beta} \tilde{\cal F}_{\dot\beta}
- a_5 \rho_\alpha E_{\dot\alpha(2)} B^{\dot\alpha(2)}
\tilde{\cal F}^\alpha \\
\delta (\Delta_1 {\cal L}_1) &=& \frac{a_1}{m} \rho_\alpha 
e^\alpha{}_{\dot\alpha} {\cal B}^{\dot\alpha\dot\beta}
\tilde{\cal F}_{\dot\beta} - a_1 \rho_\alpha e^\alpha{}_{\dot\alpha}
\Omega^{\dot\alpha\dot\beta} \tilde{\cal F}_{\dot\beta}
+ a_5 \rho_\alpha E_{\dot\alpha(2)} B^{\dot\alpha(2)}
\tilde{\cal F}^\alpha
\end{eqnarray*}
\begin{equation}
\delta ({\cal L}_1 + \Delta_1 {\cal L}_1) = \frac{a_1}{m} \rho_\alpha 
e^\alpha{}_{\dot\alpha} {\cal B}^{\dot\alpha\dot\beta}
\tilde{\cal F}_{\dot\beta} 
\end{equation}
and the only possibility to compensate these variations is
\begin{equation}
\Delta_2 = - \frac{a_1}{m^2} \phi_\alpha e^\alpha{}_{\dot\alpha}
{\cal B}^{\dot\alpha\dot\beta} \tilde{\cal F}_{\dot\beta} + h.c.
\end{equation}
The resulting cubic vertex looks like
\begin{eqnarray}
{\cal L}_1 &=& [ a_1 \Phi_{\dot\alpha} e_\alpha{}^{\dot\alpha}
\Omega^{\alpha\beta} - a_5 \Phi^\beta E_{\dot\alpha(2)}
B^{\dot\alpha(2)} - a_6 \phi_{\dot\alpha}  E_\alpha{}^{\dot\alpha}
B^{\alpha\beta}] \Psi_\beta \nonumber \\
 && - \frac{a_1}{m} \Phi_{\dot\alpha} e_\alpha{}^{\dot\alpha}
B^{\alpha\beta} \tilde{\cal F}_\beta - \frac{a_1}{m^2} \phi_\alpha
e^\alpha{}_{\dot\alpha} {\cal B}^{\dot\alpha\dot\beta} 
\tilde{\cal F}_{\dot\beta} + h.c.
\end{eqnarray} 
and in the unitary gauge we indeed have
\begin{equation}
{\cal L}_1 = a_1 \Phi_{\dot\alpha} e_\alpha{}^{\dot\alpha}
\Omega^{\alpha\beta} \Psi_\beta + h.c.
\end{equation}

\section*{Acknowledgments} 

Author is grateful to Nicolas Boulanger for stimulating discussions.

\appendix

\section{Frame-like multispinor formalism}

In the frame-like multispinor formalism we use all objects are forms
(fields are one-forms or zero-forms, gauge invariant curvatures are
two-forms or one-forms, while all the terms in the Lagrangians are
four-forms) having some number of dotted and un-dotted spinor indices
$\alpha,\dot\alpha = 1,2$. Coordinate free description of (anti) de
Sitter space is achieved with the background frame
$e^{\alpha\dot\alpha}$ and covariant external derivative $D$ (which
contains a background Lorentz connection) which satisfy
\begin{equation}
D \wedge e^{\alpha\dot\alpha} = 0, \qquad
D \wedge D \zeta^\alpha = 2\Lambda E^\alpha{}_\beta \zeta^\beta
\end{equation}
Here basic two-forms $E^{\alpha(2)}$, $E^{\dot\alpha(2)}$ are defined
as follows
\begin{equation}
e^{\alpha\dot\alpha} \wedge e^{\beta\dot\beta} =
\epsilon^{\alpha\beta} E^{\dot\alpha\dot\beta}
+ e^{\dot\alpha\dot\beta} E^{\alpha\beta}
\end{equation}
A complete basis of forms also contains three-form
$E^{\alpha\dot\alpha}$ and four-form $E$:
\begin{equation}
E^{\alpha(2)} \wedge e^{\beta\dot\alpha} =
\epsilon^{\alpha\beta} E^{\alpha\dot\alpha}, \qquad
E^{\alpha\dot\alpha} \wedge e^{\beta\dot\beta}
= \epsilon^{\alpha\beta} \epsilon^{\dot\alpha\dot\beta} E
\end{equation}
In the main text all wedge product signs are omitted.


\begin{thebibliography}{10}

\bibitem{DW01}
S.~Deser, A.~Waldron
{\it "Gauge Invariance and Phases of Massive Higher Spins in (A)dS",}
Phys. Rev. Lett. {\bf 87} (2001) 031601, arXiv:hep-th/0102166.

\bibitem{DW01a}
S.~Deser, A.~Waldron
{\it "Partial Masslessness of Higher Spins in (A)dS",}
Nucl. Phys. {\bf B607} (2001) 577, arXiv:hep-th/0103198.

\bibitem{DW01c}
S.~Deser, A.~Waldron
{\it "Null Propagation of Partially Massless Higher Spins in (A)dS and
  Cosmological Constant Speculations",}
Phys. Lett. {\bf B513} (2001) 137, arXiv:hep-th/0105181.

\bibitem{Zin01}
Yu.~M. Zinoviev
{\it "On Massive High Spin Particles in (A)dS",} arXiv:hep-th/0108192.

\bibitem{Met06}
R.~R. Metsaev
{\it "Gauge invariant formulation of massive totally symmetric
fermionic fields in (A)dS space",}
Phys. Lett. {\bf B643} (2006) 205-212, arXiv:hep-th/0609029.

\bibitem{Met22a}
R.R. Metsaev
{\it "Light-cone gauge massive and partially-massless fields in
AdS(4)",}
Phys. Lett. B {\bf 839} (2023) 137790, arXiv:2212.14728.

\bibitem{DJW12}
S.~Deser, E.~Joung, A.~Waldron
{\it "Partial Masslessness and Conformal Gravity",}
J. Phys. {\bf A46} (2013) 214019, arXiv:1208.1307.

\bibitem{DJW13}
S.~Deser, E.~Joung, A.~Waldron
{\it "Gravitational- and Self- Coupling of Partially Massless Spin
2",}
Phys. Rev. {\bf D86} (2012) 104004, arXiv:1301.4181.

\bibitem{HSS13}
S.~F. Hassan, Angnis Schmidt-May, Mikael von Strauss
{\it "Higher Derivative Gravity and Conformal Gravity From Bimetric
and Partially Massless Bimetric Theory",}
Universe {\bf 1} (2015) 092, arXiv:1303.6940.

\bibitem{Zin06}
Yu.~M. Zinoviev
{\it "On massive spin 2 interactions",}
Nucl. Phys. {\bf B770} (2007) 83-106, arXiv:hep-th/0609170.

\bibitem{Zin14}
Yu.~M. Zinoviev
{\it "Massive spin-2 in the Fradkin-Vasiliev formalism. I. Partially
massless case",}
Nucl. Phys. {\bf B886} (2014) 712, arXiv:1405.4065.

\bibitem{RR12}
Claudia de~Rham, Sebastien Renaux-Petel
{\it "Massive Gravity on de Sitter and Unique Candidate for Partially
Massless Gravity",}
JCAP {\bf 1301} (2013) 035, arXiv:1206.3482.

\bibitem{HSS12}
S.~F. Hassan, Angnis Schmidt-May, Mikael von Strauss
{\it "On Partially Massless Bimetric Gravity",}
Phys. Lett. {\bf B726} (2013) 834, arXiv:1208.1797.

\bibitem{HSS12d}
S.~F. Hassan, Angnis Schmidt-May, Mikael von Strauss
{\it "Bimetric Theory and Partial Masslessness with Lanczos-Lovelock
Terms in Arbitrary Dimensions",}
Class. Quant. Grav. {\bf 30} (2103) 184010, arXiv:1212.4525.

\bibitem{DSW13}
S.~Deser, M.~Sandora, A.~Waldron
{\it "Nonlinear Partially Massless from Massive Gravity?",}
Phys. Rev. {\bf D 87} (2013) 101501, arXiv:1301.5621.

\bibitem{RHRT13}
Claudia de~Rham, Kurt Hinterbichler, Rachel~A. Rosen, Andrew~J. Tolley
{\it "Evidence for and Obstructions to Non-Linear Partially Massless
Gravity",}
Phys. Rev. {\bf D88} (2013) 024003, arXiv:1302.0025.

\bibitem{JLT14}
Euihun Joung, Wenliang Li, M.~Taronna
{\it "No unitary theory of PM spin two and gravity",}
Phys. Rev. Lett. {\bf 113} (2014) 091101, arXiv:1406.2335.

\bibitem{JR15}
Sebastian Garcia-Saenz, Rachel~A. Rosen
{\it "A non-linear extension of the spin-2 partially massless
symmetry",}
JHEP {\bf 05} (2015) 042, arXiv:1410.8734.

\bibitem{GHJMR16}
Sebastian Garcia-Saenz, Kurt Hinterbichler, Austin Joyce, Ermis
Mitsou, Rachel~A. Rosen
{\it "No-go for Partially Massless Spin-2 Yang-Mills",}
JHEP {\bf 02} (2016) 043, arXiv:1511.03270.

\bibitem{BDGT19}
Nicolas Boulanger, Cédric Deffayet, Sebastian Garcia-Saenz, Lucas
Traina
{\it "A theory for multiple partially massless spin-2 fields",}
Phys. Rev. D {\bf 100} (2019) 101701, arXiv:1906.03868.

\bibitem{JMP19}
Euihun Joung, Karapet Mkrtchyan, Gabriel Poghosyan
{\it "Looking for partially-massless gravity",}
JHEP {\bf 07} (2019) 116, arXiv:1904.05915.

\bibitem{ST21}
Charlotte Sleight, Massimo Taronna
{\it "On the consistency of (partially-)massless matter couplings in
de Sitter space",}
JHEP {\bf 10} (2021) 156, arXiv:2106.00366.

\bibitem{BGPT24}
Nicolas Boulanger, Sebastian Garcia-Saenz, Songsong Pan, Lucas Traina
{\it "Cubic interactions for massless and partially massless spin-1
and spin-2 fields",}
JHEP {\bf 11} (2024) 019, arXiv:2407.05865.

\bibitem{GHR18}
Sebastian Garcia-Saenz, Kurt Hinterbichler, Rachel~A. Rosen
{\it "Supersymmetric Partially Massless Fields and Non-Unitary
Superconformal Representations",}
JHEP {\bf 11} (2018) 166, arXiv:1810.01881.

\bibitem{BKhSZ19a}
I.~L. Buchbinder, M.~V. Khabarov, T.~V. Snegirev, Yu.~M. Zinoviev
{\it "Lagrangian description of the partially massless higher spin N=1
  supermultiplets in $AdS_4$ space",}
JHEP {\bf 08} (2019) 116, arXiv:1904.01959.

\bibitem{BGHR21}
Noah Bittermann, Sebastian Garcia-Saenz, Kurt Hinterbichler, Rachel~A.
Rosen
{\it "${\cal N}=2$ Supersymmetric Partially Massless Fields and
Non-Unitary Superconformal Representations",}
JHEP {\bf 08} (2021) 115, arXiv:2011.05994.

\bibitem{BKhSZ19}
I.~L. Buchbinder, M.V. Khabarov, T.~V. Snegirev, Yu.~M. Zinoviev
{\it "Lagrangian formulation of the massive higher spin $N=1$
supermultiplets in $AdS_4$ space",}
Nucl. Phys. {\bf B942} (2019) 1-29, arXiv:1901.09637.

\bibitem{Zin24}
Yu.~M. Zinoviev
{\it "On massive higher spin supermultiplets in d=4",}
JHEP {\bf 10} (2024) 222, arXiv:2408.08674.

\bibitem{Zin08b}
Yu.~M. Zinoviev
{\it "Frame-like gauge invariant formulation for massive high spin
particles",}
Nucl. Phys. {\bf B808} (2009) 185, arXiv:0808.1778.

\bibitem{PV10}
D.~S. Ponomarev, M.~A. Vasiliev
{\it "Frame-Like Action and Unfolded Formulation for Massive
Higher-Spin Fields",}
Nucl. Phys. {\bf B839} (2010) 466, arXiv:1001.0062.

\bibitem{KhZ19}
M.V. Khabarov, Yu.~M. Zinoviev
{\it "Massive higher spin fields in the frame-like multispinor
formalism",}
Nucl. Phys. {\bf B948} (2019) 114773, arXiv:1906.03438.

\bibitem{BLT23}
Nicolas Boulanger, Guillaume Lhost, Sylvain Thomée
{\it "Consistent couplings between a massive spin-3/2 field and a
partially massless spin-2 field",}
Universe {\bf 9} (2023) 482, arXiv:2310.05522.

\bibitem{Zin18}
Yu.~M. Zinoviev
{\it "On massive super(bi)gravity in the constructive approach",}
Class. Quant. Grav. {\bf 35} (2018) 175006, arXiv:1805.01650.

\bibitem{Zin18a}
Yu.~M. Zinoviev
{\it "On partially massless supergravity",}
Phys. Part. Nucl. {\bf 49} (2018) 850.

\bibitem{KhZ20}
M.~V. Khabarov, Yu.~M. Zinoviev
{\it "Massive higher spin supermultiplets unfolded",}
Nucl. Phys. {\bf B953} (2020) 114959, arXiv:2001.07903.

\bibitem{BDGT18}
Nicolas Boulanger, Cedric Deffayet, Sebastian Garcia-Saenz, Lucas
Traina
{\it "Consistent deformations of free massive field theories in the
  Stueckelberg formulation",}
JHEP {\bf 07} (2018) 021, arXiv:1806.04695.

\bibitem{Zin24a}
Yu.~M. Zinoviev
{\it "On the Fradkin-Vasiliev formalism in d=4",} arXiv:2410.16798.

\end{thebibliography}
\end{document}